\documentclass[conference]{IEEEtran}
\IEEEoverridecommandlockouts
\usepackage{cite}
\usepackage{amsmath,amssymb,amsfonts}
\usepackage{algorithmic}
\usepackage{graphicx}
\usepackage{textcomp}
\usepackage{xcolor}
\usepackage{booktabs}
\usepackage{adjustbox}
\usepackage{listings}
\def\BibTeX{{\rm B\kern-.05em{\sc i\kern-.025em b}\kern-.08em
    T\kern-.1667em\lower.7ex\hbox{E}\kern-.125emX}}
\begin{document}

\title{Rtl2lean: Automated RTL-to-Lean Translation with Hierarchical Theorem Generation and Lemma Reuse
}

\author{
\IEEEauthorblockN{
Hongqin Lyu\textsuperscript{1,2},
Junxing Dong\textsuperscript{3},
Yonghao Wang\textsuperscript{1}
Zhiteng Chao\textsuperscript{1},
Tiancheng Wang\textsuperscript{1,2}
and
Huawei Li\textsuperscript{1,2}}
\IEEEauthorblockA{\textsuperscript{1}State Key Lab of Processors, Institute of Computing Technology, CAS, Beijing, China}
\IEEEauthorblockA{\textsuperscript{2}University of Chinese Academy of Sciences, Beijing, China}
\IEEEauthorblockA{\textsuperscript{3}NorthChina University of Science and Technology}
}

\maketitle

\begin{abstract}
Formal verification with interactive theorem provers can provide strong
correctness guarantees for register transfer level designs, but applying it
to existing SystemVerilog code requires substantial manual effort in semantic
modeling and proof construction. This paper presents Rtl2lean, a framework
that automatically translates RTL designs into executable Lean 4 models and
builds a hierarchical theorem library for subsequent verification. The
generated model represents hardware execution as a pure state transition
function, while a four layer theorem framework captures combinational
semantics, sequential updates, single cycle behavior, and reachability and
invariants. When a high level property cannot be discharged by the existing
theorem base, an LLM based proving loop proposes intermediate lemmas from the
current proof context and Lean feedback. Only lemmas accepted by the Lean
kernel are added to the reusable lemma pool. Experiments on six SystemVerilog
designs generate 403 theorems, all of which are successfully checked by Lean.
Among 358 foundational lemmas, 287 are available for automatic reuse, yielding
a reusable lemma ratio of 80.2 percent. The results demonstrate that Rtl2lean
can construct machine checked RTL proof libraries with low checking overhead
and substantial cross property lemma reuse.
\end{abstract}

\begin{IEEEkeywords}
RTL verification, Lean 4, theorem proving, lemma generation, large language models
\end{IEEEkeywords}

\section{Introduction}
As integrated circuits continue to grow in scale and complexity, register-transfer level (RTL) verification has become one of the most critical and costly stages in modern hardware development\cite{foster2015trends}. Processors and system-on-chip designs often contain complex datapaths, control state machines, pipelined structures, and a large number of parameterized modules. Even a minor error, such as a missing state transition, an unhandled boundary condition, or an incorrect bit-vector operation, may remain undetected until system integration or even after tape-out, leading to substantial debugging and repair costs. Therefore, establishing rigorous and scalable correctness guarantees for RTL designs at an early stage is a central challenge in hardware design automation.

Existing RTL verification flows mainly rely on dynamic simulation, model checking, equivalence checking, and formal verification based on SAT/SMT solvers. Simulation is practical and scalable for engineering use, but it can examine only a finite set of input vectors and execution traces. Model checking and bounded model checking can automatically explore the state space and have been widely used for both hardware and software verification \cite{clarke2018handbook, biere1999symbolic}. However, as the number of registers, pipeline depth, and overall state space increase, the solving process is often limited by state explosion and the rapid growth of constraints \cite{clarke2012state}.

Unlike automated model checking, which relies on exploring a finite state space, interactive theorem provers such as Lean 4 and Coq allow users to define hardware semantics in formal logic and construct machine-checked proofs that hold for arbitrary inputs and execution lengths. Lean 4 further provides dependent types, extensible metaprogramming, and efficient code generation, making it suitable both for expressing RTL designs and their properties and for building domain-specific proof automation \cite{demoura2021lean4}. Prior work such as Kami and Silver Oak has shown that embedding hardware semantics in proof assistants can support modular and parameterized verification while providing stronger correctness guarantees than finite simulation and bounded checking \cite{choi2017kami,projectoakSilverOak}. However, applying this approach to existing RTL designs remains costly. The event-driven semantics of the verilog differ substantially from Lean’s reasoning model\cite{ieee1800_2023} based on pure functions and type theory, while proofs of complex hardware properties often require manually constructed auxiliary definitions, invariants, and intermediate lemmas.

Fortunately, large language models (LLMs) have recently demonstrated strong heuristic reasoning capabilities in both code generation and formal theorem proving. Early work such as GPT-f showed that generative language models can be used to predict formal proof steps \cite{polu2020generative}. LeanDojo further introduced infrastructure for interacting with the Lean proof environment, extracting proof data, and retrieving relevant premises, and demonstrated that retrieval-based premise selection can improve automated theorem proving \cite{yang2023leandojo}. In the hardware domain, the recent CktFormalizer \cite{xiong2026cktformalizer} framework translates natural-language hardware specifications into type-safe hardware descriptions embedded in Lean 4, and uses Lean for compilation checking and functional equivalence proofs. Together, these studies suggest that LLMs can identify promising reasoning steps from the current proof state and existing proof libraries, thereby assisting theorem provers in proof search.

However, this paper focuses on a different yet complementary question: \textbf{Can existing RTL designs be automatically translated into a Lean 4 formal library, while large language models dynamically generate intermediate lemmas to improve the automation of RTL theorem proving?} Achieving this goal involves two main challenges. First, there is currently no tool that can automatically translate RTL into Lean 4 representations. Because RTL behavior is jointly determined by current inputs, register states, and combinational and sequential logic, the resulting Lean model must hide unnecessary Verilog execution details while faithfully preserving the cycle-level semantics of the original design. Second, proofs of complex RTL properties often rely on highly design-specific intermediate lemmas, such as state invariants, bit-vector properties, and multi-cycle relations. These lemmas are difficult to enumerate in advance and are also challenging for existing proof automation techniques to generate directly.

To address these challenges, we propose Rtl2lean, an LLM-assisted Lean 4 formal verification framework for RTL designs. First, we develop a lightweight RTL-to-Lean 4 compiler that translates combinational logic, state updates, and bit-vector operations into Lean definitions with explicit cycle-level semantics, providing a unified representation for formally specifying and proving RTL properties. Second, we integrate an LLM into Lean’s interactive proof loop. When a proof attempt fails, the framework collects the current proof goal, relevant definitions, and error feedback, and uses the LLM to generate candidate intermediate lemmas. Each generated lemma must be verified by the Lean kernel before it can be used to advance the proof.

The main contributions of this work are as follows:
\begin{enumerate}
    \item An automated RTL-to-Lean translation flow is proposed to convert
    combinational logic, sequential state updates, and fixed-width bit-vector
    operations into Lean definitions with explicit cycle-level semantics.
    \item An LLM-assisted theorem proving mechanism is introduced to generate
    intermediate lemmas from the current proof goal, local context, and Lean
    error feedback through an iterative generate--check--repair process.
    \item Experiments on six SystemVerilog designs demonstrate a 100 percent
    proof success rate and an 80.2 percent reusable lemma ratio.
\end{enumerate}

\section{Background}

\subsection{Interactive Theorem Proving}
The goal of theorem proving is to establish that a proposition follows from a set of formally specified assumptions through valid logical inference. Let $\Gamma$ denote a logical context containing variables, assumptions, definitions, and previously proved results, and let $P$ denote the proposition to be proved. A proof task can be written as
\begin{equation}
    \Gamma \vdash P,
\end{equation}
where $\vdash$ indicates that $P$ can be derived from the information contained in $\Gamma$ using valid inference rules. A formal proof is therefore a sequence of logically valid steps that transforms the initial goal into propositions that can be directly established from the context.

Lean~4 is an interactive theorem prover based on dependent type theory and the propositions-as-types principle. Under this principle, a proposition is represented as a type, while a proof of that proposition is represented as a term of the corresponding type. Proving a proposition $P$ is therefore equivalent to constructing a proof term $p$ such that
\begin{equation}
    \Gamma \vdash p : P.
\end{equation}
The Lean kernel checks whether the constructed term indeed has type $P$. A theorem is accepted only if its proof term passes kernel type checking. 

\subsection{Definitions, Theorems, and Intermediate Lemmas}

A formal development is primarily composed of definitions, theorems, and lemmas. Definitions specify the objects being reasoned about and their formal semantics. Theorems express the final correctness properties to be established, while lemmas capture intermediate results that can be reused during the proof.

Let $T$ be the target theorem. When directly proving $T$ is difficult, the proof can be decomposed by first establishing a set of intermediate lemmas,
\begin{equation}
    L_1, L_2, \ldots, L_n,
\end{equation}
and then using them to derive the final result:
\begin{equation}
    \Gamma \vdash L_1,\quad
    \Gamma \vdash L_2,\quad \ldots,\quad
    \Gamma \vdash L_n,
\end{equation}
\begin{equation}
    \Gamma, L_1, L_2, \ldots, L_n \vdash T.
\end{equation}
Intermediate lemmas do not alter the meaning of the target theorem. Instead, they expose auxiliary properties required by the proof and decompose a complex reasoning task into smaller logical obligations.

\begin{figure*}[ht]
  \centering
  \includegraphics[width=\linewidth]{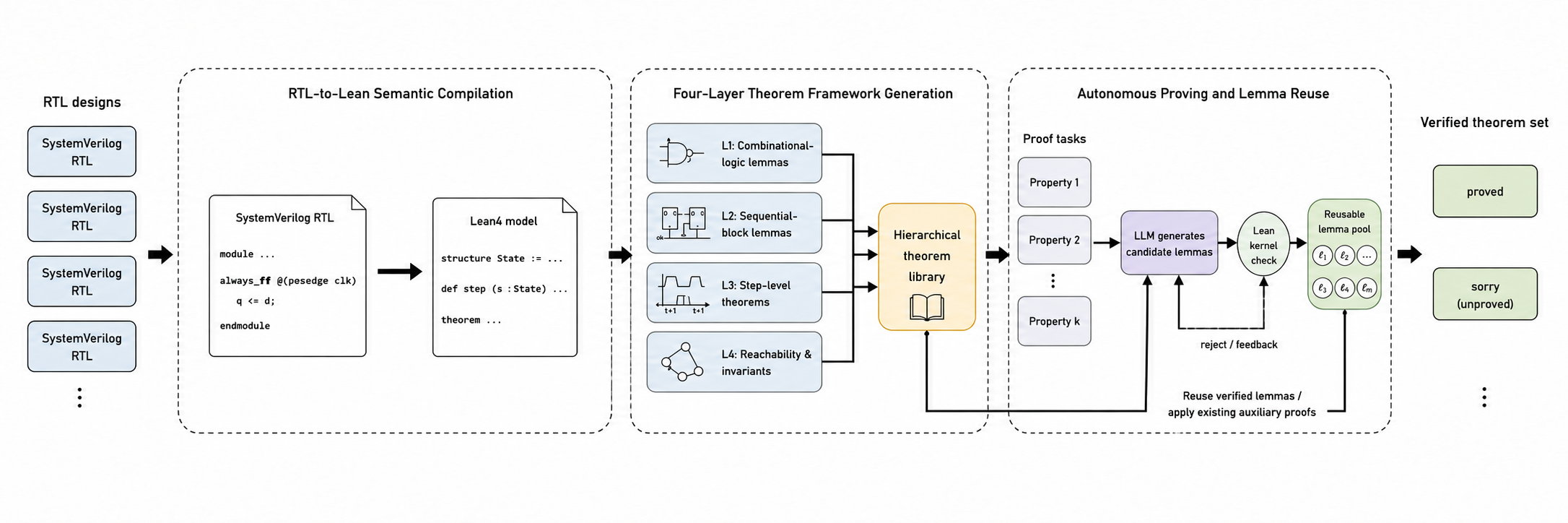}
  \caption{RTL2LEAN framework.}
  \label{fig:framework}
\end{figure*}

\section{Framework of rtl2lean}
Before presenting the technical details, we first introduce the
overall workflow of \textit{rtl2lean}, as illustrated in
Fig.~\ref{fig:framework}. The framework consists of three stages.
\begin{enumerate}
    \item \emph{RTL-to-Lean semantic compilation}:  translates
SystemVerilog RTL into a pure functional state-transition model
in Lean~4
    \item \emph{Theorem framework generation}: constructs
a four-layer proof skeleton covering combinational logic,
sequential processes, single-cycle behavior, and reachability or
invariants.
    \item \emph{Autonomous proving with lemma reuse}: proves high-level specifications through a Lean-guided feedback
loop and stores verified theorems in a lemma pool for subsequent
proof tasks.
\end{enumerate}

\subsection{RTL-to-Lean Semantic Compilation}
\label{subsec:rtl_to_lean}

SystemVerilog combines width-sensitive operations, blocking and
non-blocking assignments, conditional control flow, and distinct
combinational and sequential processes, making a direct encoding
into a theorem prover difficult to reason about. To address this
issue, \textit{rtl2lean} first parses the RTL design with PySlang
and constructs a typed abstract syntax tree. The resulting syntax
tree is then lowered into an intermediate representation, where
conditional updates are normalized as selection expressions and
different assignment types retain their corresponding update
semantics. After standard optimizations, such as constant folding,
dead-code elimination, and common-subexpression elimination, the
backend generates a Lean~4 model consisting of \texttt{State},
\texttt{Inputs}, \texttt{Outputs}, and a single-cycle transition
function \texttt{step}.

The generated model represents hardware execution as a pure
function. Given the current state \(s\) and input \(i\),
\(\operatorname{step}(s,i)\) returns the next state without side
effects. For example, the behavior of a counter with reset and
enable signals can be abstracted as
\begin{equation}
\operatorname{step}(s,i)=
\begin{cases}
\operatorname{init}, &
i.\mathrm{rst}=\mathrm{true},\\
s[\mathrm{count}\leftarrow s.\mathrm{count}+1], &
i.\mathrm{enable}=\mathrm{true},\\
s, & \text{otherwise}.
\end{cases}
\label{eq:counter_step}
\end{equation}

This functional encoding preserves the cycle-level transition
semantics of the RTL design while exposing an explicit state
relation on which subsequent theorems can be constructed.

\begin{figure}[t]
  \centering
  \includegraphics[width=\columnwidth]
    {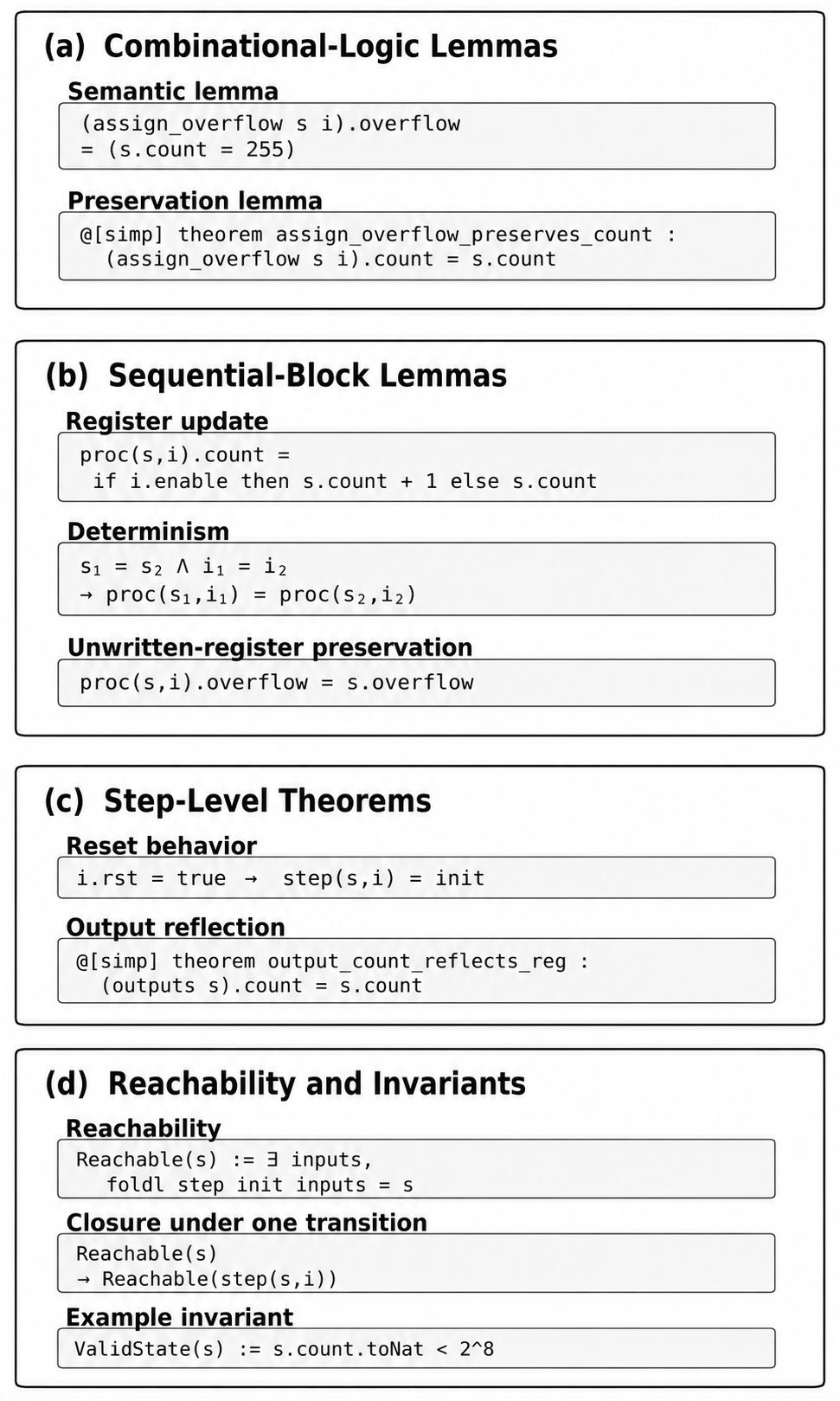}
  \caption{Representative examples generated by the four-layer.}
  \label{fig:four_layer_examples}
\end{figure}
\subsection{Four-Layer Theorem Framework Generation}
\label{subsec:four_layer_theorems}

Although RTL-to-Lean compilation establishes an executable formal semantics for
the design, directly proving high-level specifications may still require repeated
unfolding of combinational functions, sequential processes, and the single-cycle
transition function. Such proofs quickly become verbose and contain substantial
redundant reasoning. To address this issue, \textit{rtl2lean} automatically
constructs a bottom-up four-layer theorem framework from the generated Lean~4
model. Lower layers capture local hardware semantics, while higher layers reuse
these results to establish cycle-level and multi-cycle properties.

\subsubsection{Combinational-Logic Lemmas}
\label{subsubsec:comb_lemmas}

The first layer targets functions generated from continuous assignments and
combinational processes. For each combinational function, the framework
generates semantic lemmas for updated fields and preservation lemmas for
unaffected fields. As illustrated in Fig.~\ref{fig:four_layer_examples}(a),
the semantic lemma characterizes how the overflow flag is computed, while the
preservation lemma states that the counter register remains unchanged. The
preservation lemma is marked with Lean's \texttt{@[simp]} attribute, allowing
the simplifier to apply it automatically in subsequent proofs.

\subsubsection{Sequential-Block Lemmas}
\label{subsubsec:seq_lemmas}

The second layer targets state-update functions generated from sequential
processes such as \texttt{always\_ff}. For each sequential function, the
framework derives determinism lemmas, preservation lemmas for registers outside
its write set, and reset-behavior lemmas when a reset pattern is identified.
Figure~\ref{fig:four_layer_examples}(b) shows a counter update, its functional
determinism property, and preservation of an unwritten overflow register. These
lemmas package register-level behavior into reusable proof units and avoid
repeatedly unfolding the complete sequential function.

\subsubsection{Step-Level Theorems}
\label{subsubsec:step_theorems}

The third layer is built around the unified single-cycle transition function
\texttt{step}, which combines reset handling, combinational evaluation, and
sequential state updates. This layer captures module-level properties such as
reset behavior, transition determinism, output reflection, and finite-state
machine transitions. In Fig.~\ref{fig:four_layer_examples}(c), reset assertion
forces the next state to equal the initial state, while an output-reflection
lemma relates the externally observed count value to the corresponding internal
register. Step-level theorems therefore provide a direct interface for reasoning
about the complete behavior of one hardware cycle.

\subsubsection{Reachability and Invariants}
\label{subsubsec:reachability_invariants}

The first three layers characterize individual functions or a single transition
step, whereas many hardware specifications concern executions of arbitrary
length. The fourth layer therefore introduces reachability to lift the
single-cycle semantics of \texttt{step} to multi-cycle executions. As shown in
Fig.~\ref{fig:four_layer_examples}(d), a state is reachable if it can be obtained
from the initial state by folding \texttt{step} over a finite input sequence.
The framework then proves that reachability is preserved by one additional
transition and introduces state invariants over all reachable states. Such
results provide an induction basis for proving higher-level safety and
functional specifications.

\subsection{Autonomous Proving and Lemma Reuse}
\label{subsec:autonomous_proving}

The previous stages generate an executable Lean~4 model and a layered theorem
base. However, high-level functional specifications often require
design-specific intermediate results that cannot be derived by fixed proof
templates alone. To address this issue, \textit{Rtl2lean} employs a
Lean-guided autonomous proving procedure. When the existing theorem base cannot
discharge the current goal, an LLM generates candidate intermediate lemmas from
the proof context and Lean error feedback. Only candidates accepted by the Lean
kernel are added to the lemma pool and reused in subsequent proof tasks.

\subsubsection{Proof-Task Construction}
\label{subsubsec:proof_task_construction}

For each high-level specification, the system first creates a Lean theorem with
an incomplete proof and collects the relevant context, including the generated
state definitions, transition functions, four-layer theorem base, and verified
lemmas stored in the lemma pool. For a target theorem \(T\), the proof context is
represented as
\begin{equation}
\mathcal{C}(T)=
\left(
M,\,
T,\,
\mathcal{L}_{\mathrm{base}},\,
\mathcal{L}_{\mathrm{pool}}
\right),
\label{eq:proof_context}
\end{equation}
where \(M\) denotes the generated Lean~4 model,
\(\mathcal{L}_{\mathrm{base}}\) is the set of automatically generated
lower-layer lemmas, and \(\mathcal{L}_{\mathrm{pool}}\) contains reusable lemmas
proved in earlier tasks.

For example, a high-level property for the counter states that the register
remains unchanged when neither reset nor enable is asserted:
\begin{equation}
i.\mathrm{rst}=\mathrm{false}
\land
i.\mathrm{enable}=\mathrm{false}
\Longrightarrow
\operatorname{step}(s,i).\mathrm{count}=s.\mathrm{count}.
\label{eq:counter_stability_spec}
\end{equation}
The system provides this target together with the definition of
\texttt{step}, the sequential-update semantics, and relevant preservation
lemmas.

\subsubsection{Lean-Guided Proving and Lemma Reuse}
\label{subsubsec:lean_guided_proving}

Given the constructed proof task, the system first attempts to solve the target
using the existing theorem base. If the proof remains incomplete, an LLM
proposes candidate intermediate lemmas according to the current goal, relevant
definitions, available lemmas, and Lean error messages. Each candidate is
checked by the Lean kernel before use. Rejected candidates are returned to the
LLM with diagnostic feedback, whereas accepted lemmas are added to the lemma
pool and used to continue the proof. The iterative process is expressed as
\begin{equation}
p_{k+1}
=
\operatorname{Generate}
\bigl(\mathcal{C}(T),e_k\bigr),
\label{eq:proof_generation}
\end{equation}
where \(p_k\) is the \(k\)-th candidate proof and \(e_k\) is the error returned
by Lean. The loop terminates when
\begin{equation}
\operatorname{Check}(T,p_k)=\mathrm{Pass}.
\label{eq:proof_check}
\end{equation}

After a successful proof, the lemma pool is updated as
\begin{equation}
\mathcal{L}_{k+1}
=
\mathcal{L}_{k}
\cup
\{T\}
\cup
\mathcal{H},
\label{eq:lemma_pool_update}
\end{equation}
where \(\mathcal{H}\) denotes auxiliary lemmas or local
\texttt{have} statements discovered during proof generation. Later proofs can
reuse these results through tactics such as \texttt{rw}, \texttt{apply}, and
\texttt{simp}, avoiding repeated unfolding of the underlying RTL semantics.

For the counter property in Eq.~\eqref{eq:counter_stability_spec}, a valid proof
can be written as
\begin{lstlisting}[basicstyle=\ttfamily\footnotesize,breaklines=true]
theorem count_stable_when_disabled :
  forall (s : counterState) (i : counterInputs),
    i.rst = false ->
    i.enable = false ->
    (step s i).count = s.count := by
  intro s i hrst hen
  simp [step, proc_alwaysff, hrst, hen]
\end{lstlisting}
Once verified, this theorem can be reused to prove related properties, such as
the stability of the corresponding output value.

\section{Expermental}
\subsection{Expermental set}
To evaluate the effectiveness of Rtl2lean in RTL formal modeling, theorem generation, and Lean-based machine checking, we select six SystemVerilog designs with diverse functional and structural characteristics from OpenCores\cite{opencores}, as shown in Table~\ref{tab:benchmarks}. This benchmark suite enables the applicability of Rtl2lean to be evaluated across different design types and hardware structures. To validate the effectiveness of the proposed method, Lean~4.26\cite{lean4260} is used to check the Lean models, foundational lemmas, and high-level properties generated by Rtl2lean. All experiments are conducted on a server equipped with an Intel(R) Xeon(R) Gold 6148 CPU and 629~GB of RAM.

\begin{table}[t]
    \centering
    \caption{OpenCores RTL designs used in the experiments}
    \label{tab:benchmarks}
    \resizebox{\columnwidth}{!}{
    \begin{tabular}{ll}
        \toprule
        \textbf{Design} & \textbf{Main Function} \\
        \midrule
        \texttt{debounce\_edge}
        & Debouncing and edge detection \\

        \texttt{fir\_filter}
        & Finite impulse response filtering \\

        \texttt{pwm\_generator}
        & Pulse-width modulation \\

        \texttt{lfsr\_crc}
        & LFSR and CRC computation\\

        \texttt{sync\_fifo}
        & Synchronous FIFO buffering\\

        \texttt{ps2\_keyboard}
        & PS/2 keyboard control \\
        \bottomrule
    \end{tabular}}
\end{table}

\subsection{Evaluation Metrics}

To evaluate the proof-generation capability, correctness, execution efficiency, and lemma reusability of Rtl2lean, we adopt the metrics summarized in Table~\ref{tab:evaluation_metrics}. The total number of theorems, denoted by $N_{\mathrm{total}}$, measures the overall scale of the formal verification artifacts generated and checked by Lean. The number of high-level properties, denoted by $N_{\mathrm{property}}$, indicates the number of RTL functional and state-related properties verified by the framework. The number of foundational lemmas, denoted by $N_{\mathrm{lemma}}$, characterizes the size of the supporting proof library automatically generated for proving these high-level properties.

The proof success rate measures the proportion of generated theorems that are successfully accepted by Lean and is defined as
\begin{equation}
    \text{Proof Success Rate}
    =
    \frac{N_{\mathrm{proved}}}{N_{\mathrm{total}}}
    \times 100\%,
\end{equation}
where $N_{\mathrm{proved}}$ denotes the number of theorems successfully checked by Lean, and $N_{\mathrm{total}}$ denotes the total number of generated theorems. We additionally report the numbers of errors, unresolved goals, and proofs containing \texttt{sorry} to identify incomplete verification results.

For efficiency evaluation, we report the end-to-end execution time, denoted by $T_{\mathrm{e2e}}$, and the Lean checking time, denoted by $T_{\mathrm{Lean}}$. The former measures the total time for RTL translation, theorem and lemma generation, proof construction, and result collection, while the latter measures the time required by Lean to check the generated proof files. We also evaluate the automatic reusability of foundational lemmas by defining the reusable lemma ratio as
\begin{equation}
    R_{\mathrm{reuse}}
    =
    \frac{N_{\mathrm{reusable}}}{N_{\mathrm{lemma}}}
    \times 100\%,
\end{equation}
where $N_{\mathrm{reusable}}$ denotes the number of foundational lemmas that can be automatically applied in subsequent proofs, and $N_{\mathrm{lemma}}$ denotes the total number of foundational lemmas.

\begin{table}[t]
    \centering
    \caption{Evaluation metrics for Rtl2lean}
    \label{tab:evaluation_metrics}
    \begin{adjustbox}{max width=\columnwidth}
    \begin{tabular}{ll}
        \toprule
        \textbf{Metric} & \textbf{Symbol} \\
        \midrule
        Total theorems          & $N_{\mathrm{total}}$ \\
        High-level properties   & $N_{\mathrm{property}}$ \\
        Foundational lemmas     & $N_{\mathrm{lemma}}$ \\
        Proof success rate      & $R_{\mathrm{proof}}$ \\
        End-to-end time         & $T_{\mathrm{e2e}}$ \\
        Lean checking time      & $T_{\mathrm{Lean}}$ \\
        Reusable lemma ratio    & $R_{\mathrm{reuse}}$ \\
        \bottomrule
    \end{tabular}
    \end{adjustbox}
\end{table}

\subsection{Proof Results and Lemma Reuse Analysis}

Table~\ref{tab:proof_results} presents the proof results and lemma reuse statistics of Rtl2lean on the six RTL designs. Rtl2lean generates and verifies a total of 403 theorems, including 45 high-level properties and 358 foundational lemmas. Among the foundational lemmas, 287 can be automatically reused in subsequent proofs. All generated theorems are successfully checked by Lean, resulting in a proof success rate of 100\% for every design.

\begin{table*}[t]
    \centering
    \caption{Proof results and lemma reuse statistics of Rtl2lean}
    \label{tab:proof_results}
    \small
    \setlength{\tabcolsep}{7pt}
    \begin{tabular}{lrrrrrr}
        \toprule
        \textbf{Design}
        & $N_{\mathrm{property}}$
        & $N_{\mathrm{lemma}}$
        & $N_{\mathrm{reusable}}$
        & $N_{\mathrm{total}}$
        & $R_{\mathrm{proof}}$
        & $R_{\mathrm{reuse}}$ \\
        \midrule
        \texttt{debounce\_edge}
        & 8 & 71 & 56 & 79 & 100\% & 78.9\% \\

        \texttt{fir\_filter}
        & 7 & 22 & 14 & 29 & 100\% & 63.6\% \\

        \texttt{pwm\_generator}
        & 7 & 74 & 62 & 81 & 100\% & 83.8\% \\

        \texttt{lfsr\_crc}
        & 7 & 35 & 27 & 42 & 100\% & 77.1\% \\

        \texttt{sync\_fifo}
        & 8 & 86 & 70 & 94 & 100\% & 81.4\% \\

        \texttt{ps2\_keyboard}
        & 8 & 70 & 58 & 78 & 100\% & 82.9\% \\
        \midrule
        \textbf{Total}
        & \textbf{45}
        & \textbf{358}
        & \textbf{287}
        & \textbf{403}
        & \textbf{100\%}
        & \textbf{80.2\%} \\
        \bottomrule
    \end{tabular}
\end{table*}

As shown in Table~\ref{tab:proof_results}, the number of high-level properties remains relatively stable across the evaluated designs, ranging from seven to eight, whereas the number of foundational lemmas varies considerably, from 22 for \texttt{fir\_filter} to 86 for \texttt{sync\_fifo}. This difference indicates that the size of the foundational proof library is mainly affected by the complexity of state variables, assignment operations, and state-update dependencies, rather than solely by the number of high-level properties. In particular, \texttt{sync\_fifo} contains more state and control relations and therefore produces the largest number of foundational lemmas and total theorems.

All six designs achieve a proof success rate of 100\%, demonstrating that the Lean models and proofs generated by Rtl2lean can be successfully machine-checked across designs with different functions and structures. The overall reusable lemma ratio reaches 80.2\%, with five designs exceeding 77\%, indicating that most foundational lemmas can be reused in subsequent high-level proofs to reduce repeated reasoning over low-level assignment and state-update semantics. This result also validates the effectiveness of Rtl2lean's hierarchical proof structure, in which foundational semantic relations are proved once and then shared across multiple properties. Note that $R_{\mathrm{reuse}}$ measures the proportion of lemmas available for automatic reuse rather than their exact dynamic usage in individual proofs.

\subsection{Time Overhead}

We measure the end-to-end execution time, $T_{\mathrm{e2e}}$, and the Lean checking time, $T_{\mathrm{Lean}}$. The former includes RTL translation, theorem and lemma generation, proof construction, and result collection, while the latter measures the time required by Lean to check the generated proof files. Table~\ref{tab:time_overhead} reports the checking time for each design.

\begin{table}[t]
    \centering
    \caption{Time overhead of Rtl2lean}
    \label{tab:time_overhead}
    \small
    \setlength{\tabcolsep}{5pt}
    \begin{tabular}{lrr}
        \toprule
        \textbf{Design}
        & $N_{\mathrm{total}}$
        & $T_{\mathrm{Lean}}$ (s) \\
        \midrule
        \texttt{debounce\_edge} & 79 & 1.113 \\
        \texttt{fir\_filter}    & 29 & 0.558 \\
        \texttt{pwm\_generator} & 81 & 0.918 \\
        \texttt{lfsr\_crc}      & 42 & 0.721 \\
        \texttt{sync\_fifo}     & 94 & 0.936 \\
        \texttt{ps2\_keyboard}  & 78 & 1.327 \\
        \midrule
        \textbf{Total}          & \textbf{403} & \textbf{5.573} \\
        \bottomrule
    \end{tabular}
\end{table}

The complete proof-generation pipeline takes only $T_{\mathrm{e2e}}=0.650$~s for all six designs, since the deterministic proof templates require neither proof search nor external model calls. In comparison, Lean checking takes 5.573~s in total, indicating that most of the time is spent on definition reduction and proof-term checking. Each design is checked in less than 1.4~s, including \texttt{sync\_fifo}, which contains the largest number of theorems. However, $T_{\mathrm{Lean}}$ is not strictly proportional to $N_{\mathrm{total}}$; for example, \texttt{ps2\_keyboard} requires more checking time than \texttt{sync\_fifo} despite containing fewer theorems. This suggests that checking cost is also influenced by the complexity of state-transition logic and bit-vector expressions.

\section{Conclusion}
\label{sec:conclusion}

This paper presents Rtl2lean, a framework that translates SystemVerilog RTL
into executable Lean~4 models, generates a hierarchical theorem library, and
supports high-level verification through LLM-assisted lemma generation and
Lean-checked lemma reuse. Experiments on six RTL designs produce 403
machine-checked theorems with a 100 percent proof success rate and an
80.2 percent reusable lemma ratio. These results demonstrate the potential of
combining automated RTL translation, structured theorem generation, and
kernel-checked LLM assistance for scalable hardware verification.

\bibliographystyle{IEEEtran}
\bibliography{refs}

\vspace{12pt}

\end{document}